\documentclass[preprint,showpacs]{revtex4-1}

\usepackage{graphicx}
\usepackage{bm}
\usepackage[english]{babel}
\usepackage{dcolumn}
\usepackage{amssymb}

\begin{document}

\title{Numerical tests of the envelope theory for few-boson systems}

\author{Claude Semay}

%\authorrunning{Short form of author list} % if too long for running head

\email[E-mail: ]{claude.semay@umons.ac.be}
\affiliation{Service de Physique Nucl\'{e}aire et Subnucl\'{e}aire,
Universit\'{e} de Mons,
UMONS Research Institute for Complex Systems,
Place du Parc 20, 7000 Mons, Belgium}
\date{\today}

\date{\today}

\begin{abstract}
The envelope theory, also known as the auxiliary field method, is a simple technique to compute approximate solutions of Hamiltonians for $N$ identical particles in $D$-dimension. The accuracy of this method is tested by computing the ground state of $N$ identical bosons for various systems. A method is proposed to improve the quality of the approximations by modifying the characteristic global quantum number of the method. 
\end{abstract}
%\keywords{Bound states \and Many-body systems}
%\PACS{03.65.Ge}
%03.65.Ge Solutions of wave equations: bound states
% \subclass{MSC code1 \and MSC code2 \and more}

\maketitle

\section{Introduction}
\label{sec:intro}

The envelope theory (ET), aslo known as the auxiliary field method, is a technique to obtain approximate solutions of $N$-body Hamiltonians in quantum mechanics \cite{hall80,hall04}. It has been recently extended to treat systems with arbitrary kinematics in $D$-dimension \cite{sema13a}. The basic idea is to replace the Hamiltonian $H$ under study by an auxiliary Hamiltonian $\widetilde H$ which is solvable, the eigenvalues of $\widetilde H$ being optimized to be as close as possible to those of $H$. The method is easy to implement since it reduces to find the solution of a transcendental equation. In the most favourable cases, the approximate eigenvalue is an analytical lower or upper bound. In less favourable situations, a nonvariational numerical approximation can be computed. This is often interesting for $N$-body problems which are always difficult and very heavy to solve numerically. 

In a recent paper \cite{horn14}, accurate ground state energies and correlation functions of bosons interacting via various potentials are calculated using Explicitly Correlated Gaussian (ECG) basis \cite{suzu98}. Up to 8 particles are considered, allowing a plenty of numerical results to test the accuracy of the ET for eigenvalues and eigenvectors (the quality of eigenvectors for the ET has only been tested for $N=2$ \cite{sema10}). It is worth mentioning that the purpose of the ET is not to compete with methods such as the ECG. The main goal of the ET is to produce without great pain reliable estimations for the energy of a $N$-body system. If a lower or an upper bound can be produced, this must be sufficient for some applications or can be used as a test for heavy numerical computations. In the peculiar situations where an analytical expression is obtained, the result can give valuable insights about the system. 

In several cases, it has been shown that the approximation given by the ET can be largely improved by modifying the structure of the characteristic global quantum number of the method \cite{silv10,silv12}. The problem is that the kind of modification must be guessed for each system, and can be found precisely only by comparing with several exact (or accurate estimation of) eigenvalues. Clearly, a better defined procedure would be preferable. A universal effective quantum number for centrally symmetric 2-body systems is proposed in \cite{loba09}. We will show that this notion can be used to improve the ET. Let us note that the variational character of the approximation cannot be guaranteed when the global quantum number is modified. 

The paper is organized as follows. The ET method is recalled in Sec.~\ref{sec:et}, where a modification of the global quantum number is proposed to improve the quality of the approximation. The accuracy of the method is tested with various systems in Sec.~\ref{sec:res}. Some concluding remarks are finally given. 

\section{The envelope theory}
\label{sec:et}

Here we consider the $N$-body Hamiltonian, in a $D$ dimensional space ($D \ge 2$), for identical particles with a kinetic energy $T$, interacting via the one-body $U$ and two-body $V$ interactions ($\hbar=c=1$)
\begin{equation}
\label{HNb}
H=\sum_{i=1}^N T(|\bm p_i|) + \sum_{i=1}^N U\left(|\bm r_i - \bm R|\right) + \sum_{i\le j=1}^N V\left(|\bm r_i - \bm r_j|\right),
\end{equation}
where $\sum_{i=1}^N \bm p_i = \bm 0$ and $\bm R = \frac{1}{N}\sum_{i=1}^N \bm r_i$ is the centre of mass position. It is shown in \cite{sema13a} that an approximate eigenvalue $E$ is given by the following set of equations for a completely (anti)symmetrized state:
\begin{eqnarray}
\label{AFM1N}
&&E=N\, T(p_0) + N\, U \left( \frac{r_0}{N} \right) + C_N\, V \left( \frac{r_0}{\sqrt{C_N}} \right), \\
\label{AFM2N}
&&r_0\, p_0=Q, \\
\label{AFM3N}
&&N\, p_0\, T'(p_0) =  r_0\, U' \left( \frac{r_0}{N} \right) + \sqrt{C_N}\, r_0\, V' \left( \frac{r_0}{\sqrt{C_N}} \right),
\end{eqnarray}
where $A'(x)=dA/dx$, $C_N=N(N-1)/2$ is the number of particle pairs, and 
\begin{equation}
\label{QN}
Q = \sum_{i=1}^{N-1} (2 n_i + l_i) + (N - 1)\frac{D}{2}
\end{equation}
is a global quantum number (corresponding to $N-1$ identical harmonic oscillators). Following the forms of $T$, $U$ and $V$, the approximate value $E$ can have a variational character. The corresponding approximate eigenstate is given by \cite{silv12}
\begin{equation}\label{piphi}
\psi=\phi_{\textrm{cm}}(\bm R)\,\prod^{N-1}_{i=1}\, \varphi_{n_i l_i}(\lambda_i\, \bm x_i),
\end{equation}
where $\varphi_{n_i l_i}(\lambda_i\, \bm x_i)$ is a $D$-dimensional harmonic oscillator wavefunction \cite{yane94}, depending on the Jacobi coordinate $\bm x_i$ \cite{silv10}, and decreasing asymptotically like ${\rm e}^{-\lambda^2_i\, \bm x_i^{\, 2}/2}$ (the magnetic quantum number is omitted). The function $\phi_{\textrm{cm}}(\bm R)$, which can be an oscillator state, describes the centre of mass motion ($\bm x_N = \bm R$). The scale parameters $\lambda_i$ are given by
\begin{equation}\label{laj}
\lambda_i=\sqrt{\frac{i}{i+1}N Q}\,\frac{1}{r_0}=\sqrt{\frac{i}{i+1}\frac{N}{Q}}\,p_0.
\end{equation}
A state~(\ref{piphi}) has neither a defined total angular momentum nor a good symmetry, but its is characterized by a parity $(-1)^{Q-(N-1)D/2}$. By combining such states with the same value of $Q$, it is generally possible to build a physical state with good quantum numbers and good symmetry properties, but the task can be technically very complicated, even for $N=3$ \cite{silv85}.

It is possible to compute mean values for the power of interparticle distance $\langle r^k \rangle$ with the pair correlation function defined by
%\footnote{The factor $2/N(N-1)$ necessary to insure a correct normalization of $C(\bm r)$ is missing in \cite{horn14}.} 
\begin{equation}
\label{corr}
C(\bm r)=\frac{2}{N(N-1)} \left\langle \psi \left| \sum_{i<j=1}^N \delta(\bm r_i - \bm r_j -\bm r) \right| \psi \right\rangle.
\end{equation}
The value at origin, $\delta$, of the radial part of this density is also interesting to compute. For the bosonic ground state ($n_i=l_i=0, \ \forall i$), which is completely symmetrical, we have 
\begin{equation}
\label{corr2}
C(\bm r)= \frac{\lambda_1^D}{\pi^{D/2}}e^{-\lambda_1^2 r^2} \quad \textrm{and} \quad \delta=\frac{2\, \lambda_1^D}{\Gamma(D/2)}.
\end{equation}

In \cite{loba09}, it is shown that an effective quantum number $q$ that determines with high accuracy the level ordering of centrally symmetric 2-body systems has the following structure\footnote{For practical purposes, our definition of $\phi$ differs from the one used in \cite{loba09}.}
\begin{equation}
\label{nqnum}
q=\phi \left( n+\frac{1}{2} \right) +l+\frac{D-2}{2},
\end{equation}
where the number $\phi$ depends on the system. For instance, exact results are obtained for a harmonic oscillator with $\phi=2$, and for a Coulomb system with $\phi=1$. In the dominantly orbital state (DOS) method \cite{olss97,sema13b}, the approximate solutions are found by quantizing the radial motion around a semiclassical solution for a circular motion. The same separation in a radial quantum number $n+\frac{1}{2}$ and an orbital quantum number $l+\frac{D-2}{2}$ naturally occurs in this method. Its extension to 3-body systems seems to preserve the same separation \cite{sema13b}. Moreover, it has been shown, in the framework of the ET, that relations exist linking the energy of a $N$-body system to the energy of the corresponding 2-body system by a rescaling of the global quantum numbers \cite{silv11}. So, it is tempting to try to modify the global quantum number $Q$ as 
\begin{equation}
\label{QNphi}
Q_\phi = \sum_{i=1}^{N-1} (\phi\, n_i + l_i) + (N - 1)\frac{D+\phi-2}{2},
\end{equation}
which partly breaks the strong degeneracy of $Q$. The parameter $\phi$ can be determined by theoretical considerations, or by a fit on a single known accurate solution. In the following, $Q$ will be systematically replaced by $Q_\phi$. The genuine ET, with its possible variational solutions, is then recovered with $\phi=2$. For other values of $\phi$, the variational character of the solution cannot be guaranteed.

\section{Results}
\label{sec:res}

In this section, we consider various systems of $N$ identical particles with the same mass $m$. The ET formalism developed is valid for bosons or fermions with any values of $D$, but we will show only numerical results for bosons in the $D=3$ space. For the three first nonrelativistic systems taken from and named as in \cite{horn14}, only the ground state is computed. The fourth case is an ultrarelativistic 3-body system studied in \cite{silv10}, for which excited states are also available.

\subsection{Weakly interacting bosons}
\label{sec:wib}

In the first case, the nonrelativistic particles interact via a soft Gaussian potential
\begin{equation}\label{Hwib}
T(p)=\frac{p^2}{2 m}, \ U(s)=0, \ V(r)=-V_0\,e^{-r^2/R^2}.
\end{equation}
The resolution of the system (\ref{AFM1N}-\ref{AFM3N}) gives the following upper bound for the energy $E$ 
\begin{eqnarray}
\label{Hwibr0}
r_0&=& N^{1/2}(N-1)^{1/2}\, \sqrt{-W_0(Y)}\, R,\\
\label{HwibE}
E&=&-\frac{N(N-1)}{2}\, V_0\, Y^2\, \frac{1+2\, W_0(Y)}{W_0(Y)^2}\\
\label{HwibY}
&\textrm{with}& Y=-\frac{1}{N^{1/2}\, (N-1)}\frac{Q_\phi}{R\, \sqrt{2\,m\, V_0}}.
\end{eqnarray}
The multivalued Lambert function $W(z)$ is the inverse function of $z\,e^z$ \cite{corl96}. $W_0(z)$ is the branch defined for $z \ge -1/e$.

\begin{figure*}[htb]
\includegraphics[width=0.48\textwidth]{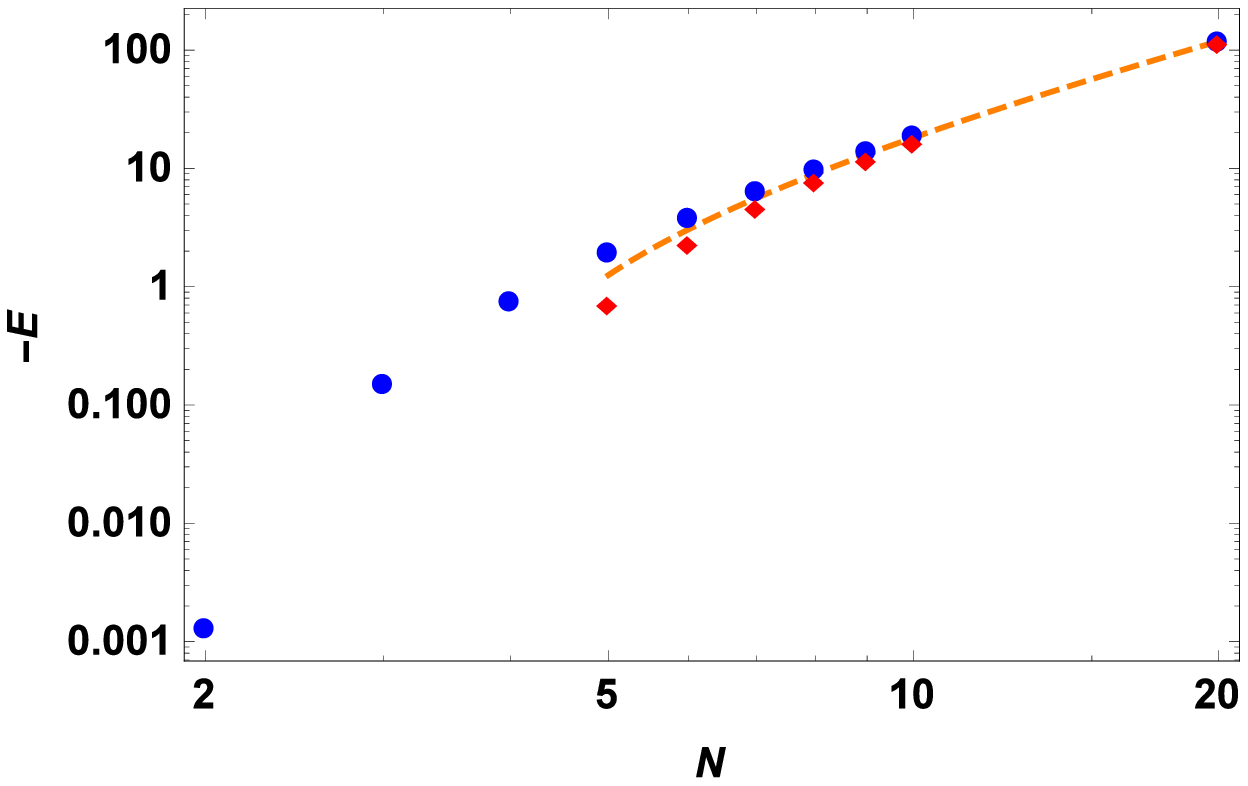} \quad
\includegraphics[width=0.48\textwidth]{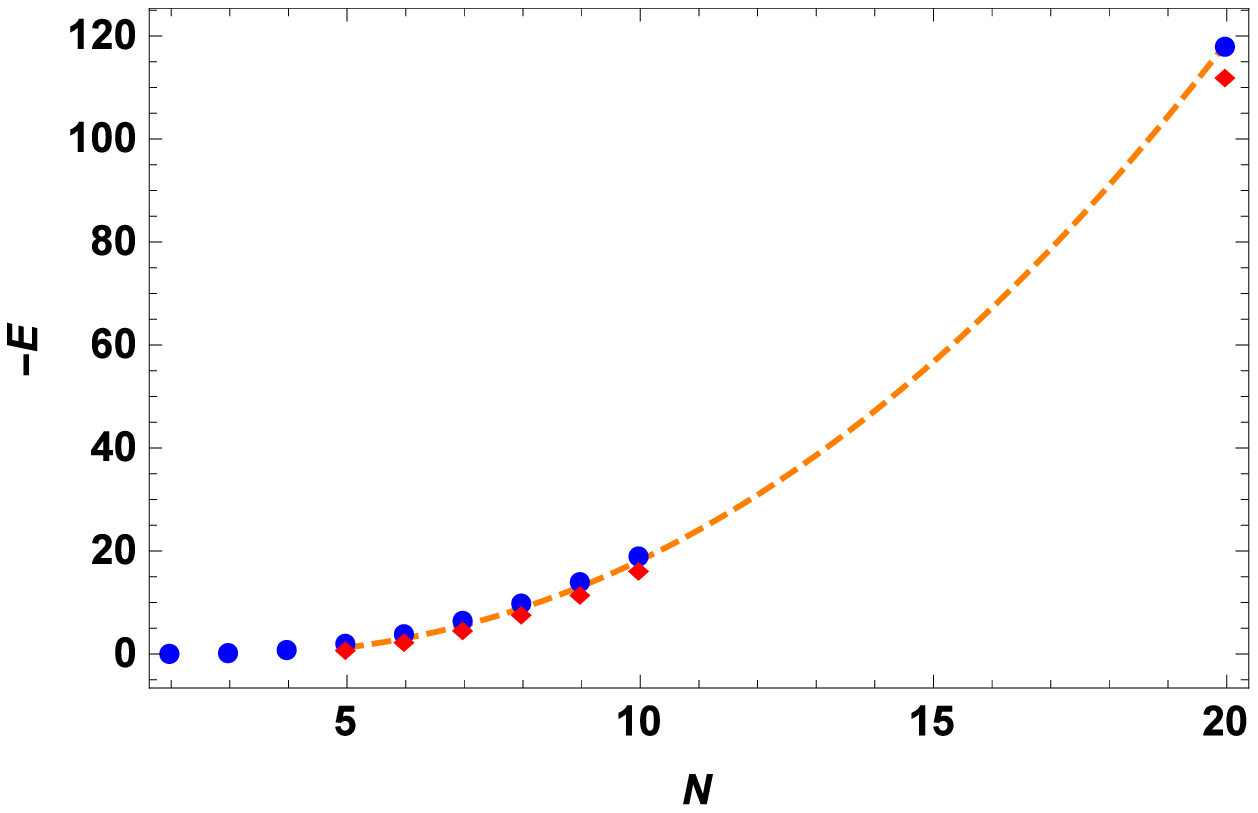}
\caption{Binding energies, $-E$, of weakly-interacting bosons: exact results (circle) \cite{horn14,timo12}, ET results for $\phi=2$ (diamond), ET results for $\phi=1.82$ (dashed line). Left: log-log plot, right: linear plot.}
\label{fig:wibE} 
\end{figure*}

In order to test the quality of the approximation with accurate data, we compare with the results given in \cite{horn14}, for the parameters $1/m=43.281307$ (a.u.)$^2$~K, $V_0=1.227$~K and $R=10.03$~a.u. Results for the ground state energies $E$ are presented in Fig.~\ref{fig:wibE} (the exact result for $N=20$ is taken from \cite{timo12}). The agreement is quite good for the genuine ET, but irrelevant values for the energies (complex or positive real numbers) are computed when the binding is very weak. This phenomenon has already been observed in \cite{silv09} for $N=2$. Note that excited values of the energies are available up to $N=6$ in \cite{gatt11}, but the ET method can only produce one relevant value of the energy for $N=6$. So no tendency can be shown as a function of $N$. A better agreement with the exact values for $N\ge 5$ is obtained for $\phi=1.82$, but the improvement is not spectacular. 

\begin{figure*}[htb]
\includegraphics[width=0.48\textwidth]{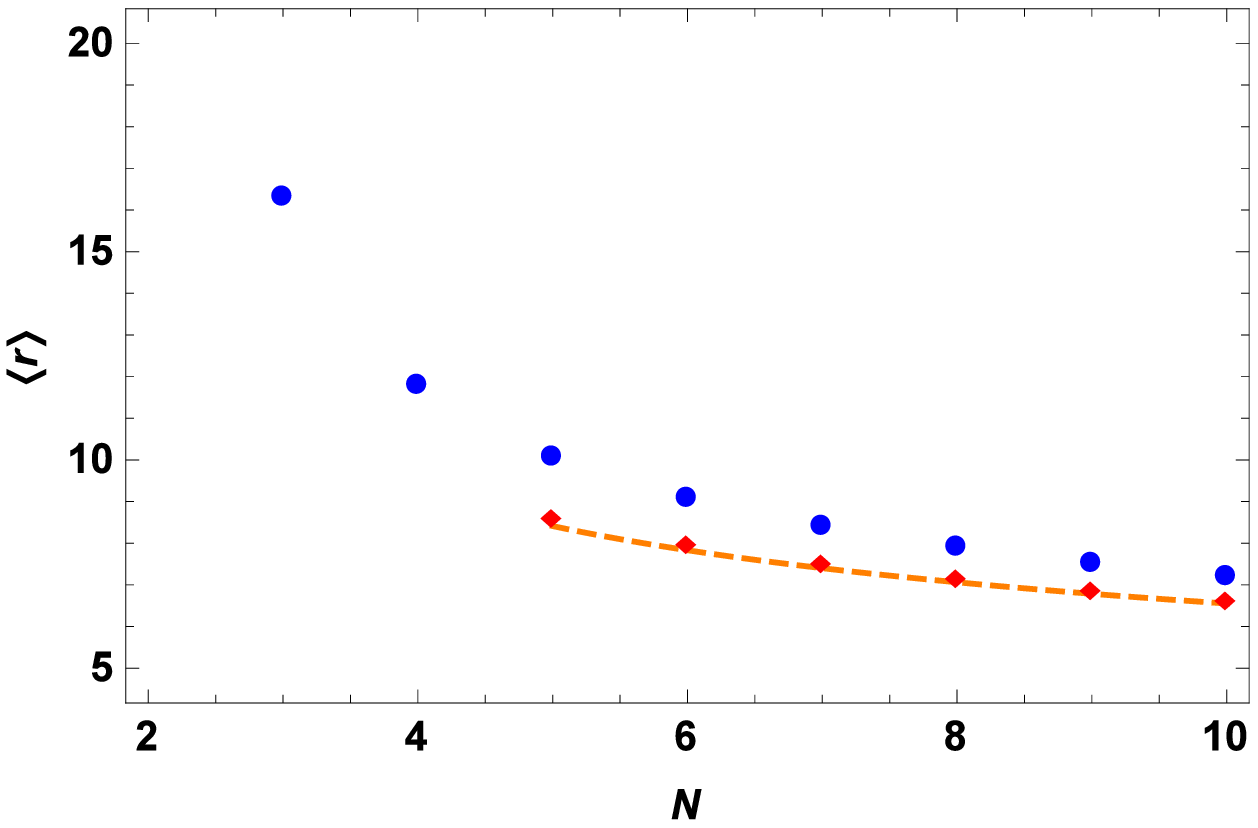} \quad
\includegraphics[width=0.48\textwidth]{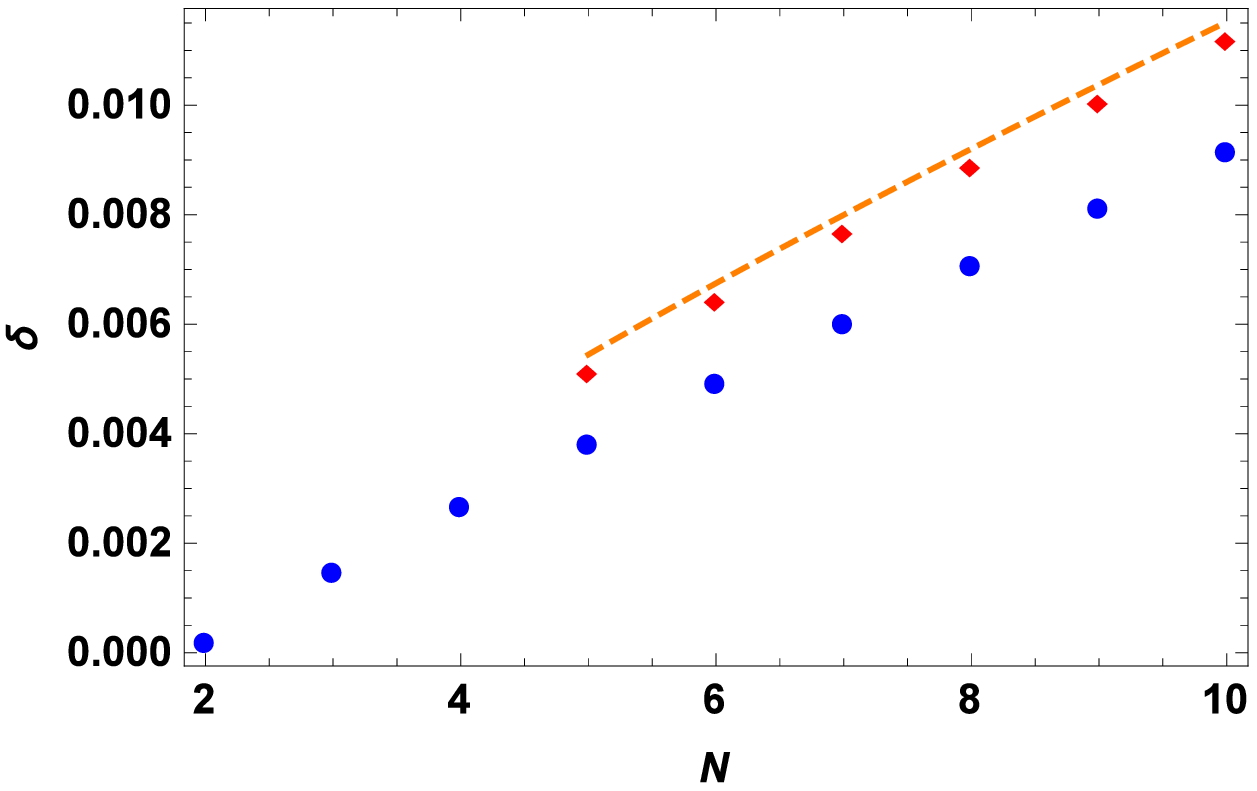}
\caption{$\langle r \rangle$ (left) and $\delta$ (right) for weakly-interacting bosons: exact results (circle) \cite{horn14}, ET results for $\phi=2$ (diamond), ET results for $\phi=1.82$ (dashed line). }
\label{fig:wibobs} 
\end{figure*}

Mean values $\langle r \rangle$ are presented with $\delta$ in Fig.~\ref{fig:wibobs}. The global agreement is good, but the value $\phi=1.82$ does not necessarily produce an improvement. Note that the values of $\delta$ are not so well reproduced. This is typical for a given value of a density. Better results are obtained for mean values, for which an integration is performed with the density over the whole domain of distances. 

\subsection{Self-gravitating bosons} 
\label{sec:sgb}

In this case, we consider nonrelativistic particle interacting via a Coulomb-like potential
\begin{equation}\label{Hsgb}
T(p)=\frac{p^2}{2 m}, \ U(s)=0, \ V(r)=-\frac{g}{r}.
\end{equation}
For a self-gravitating system, $g=m^2 G$ where $G$ is the gravitational constant. The ET method gives the following upper bound
\begin{eqnarray}
\label{Hsgbr0}
r_0&=& \frac{2^{3/2}}{N^{1/2}(N-1)^{3/2}} \frac{Q_\phi^2}{m\, g},\\
\label{HsgbE}
E&=& - \frac{N^2(N-1)^3}{16} \frac{m\, g^2}{Q_\phi^2}.
\end{eqnarray}
For $N=2$, the exact result is recovered with $\phi=1$ \cite{yane94}. Moreover, for $D=3$ and $\phi=1$, the ground state bosonic energy is
\begin{equation}
\label{Esgb_gs}
E_{\textrm{gs}} \approx -0.0625\, N^2(N-1) \, m\, g^2.
\end{equation}
This is comparable with the lower bound obtained in \cite{basd90}
\begin{equation}
\label{Esgb_gs2}
E_{\textrm{gs}} > -0.0593\, N^2(N-1) \, m\, g^2.
\end{equation}

\begin{figure*}[htb]
\includegraphics[width=0.48\textwidth]{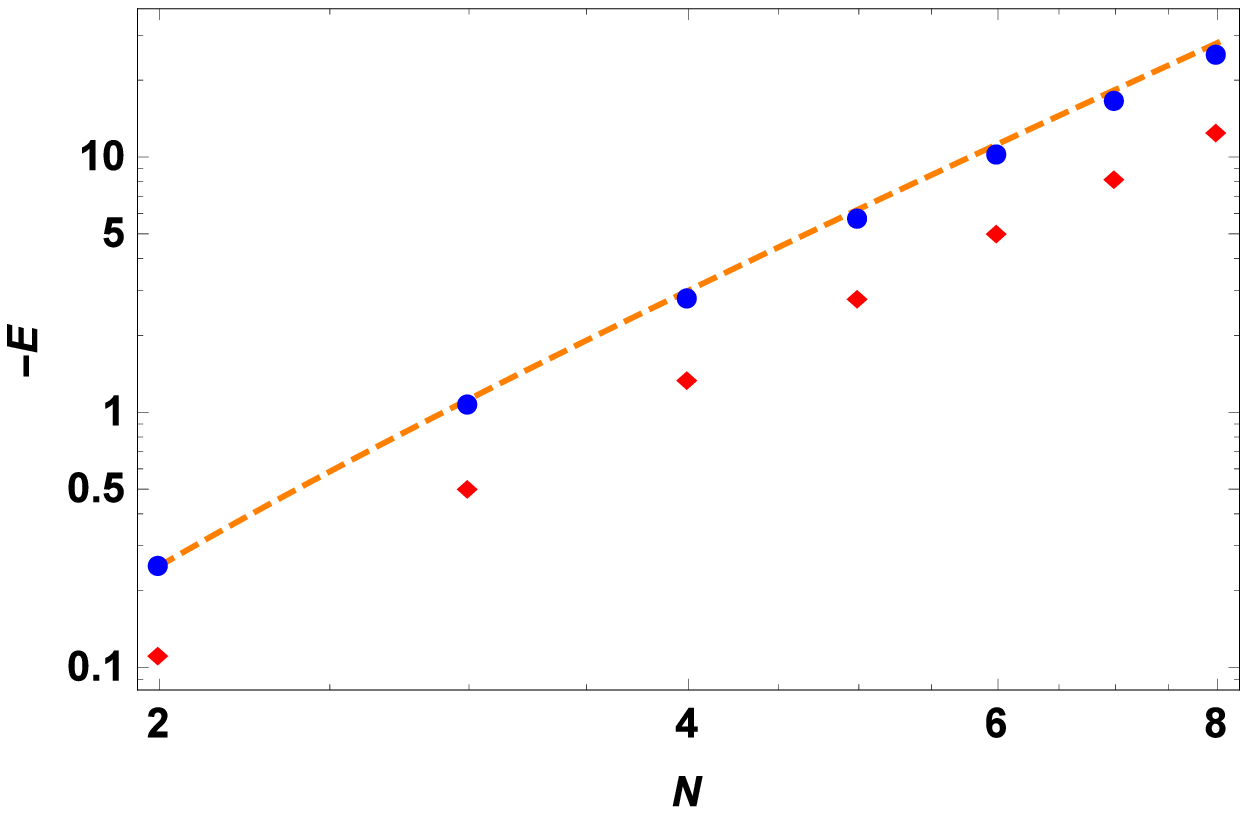} \quad
\includegraphics[width=0.48\textwidth]{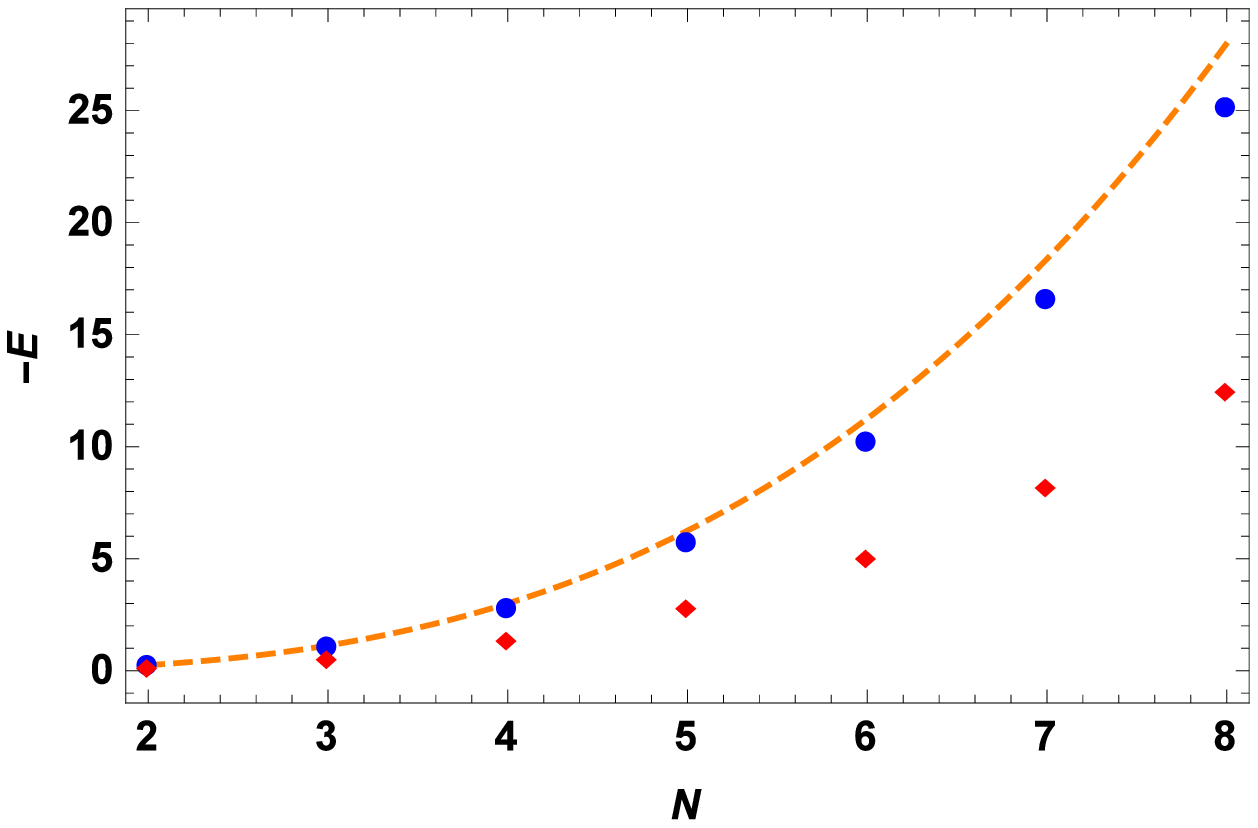}
\caption{Binding energies, $-E$, of self-gravitating bosons: exact results (circle) \cite{horn14}, ET results for $\phi=2$ (diamond), ET results for $\phi=1$ (dashed line). Left: log-log plot, right: linear plot.}
\label{fig:sgbE} 
\end{figure*}

As in the previous case, the comparison is performed with \cite{horn14}, for the parameters $m=1$ and $g=1$. Results for the ground state energies $E$ are presented in Fig.~\ref{fig:sgbE} for $\phi=2$ and $\phi=1$. If the global tendency is good for the genuine ET, as indicated by the log-log plot, a spectacular improvement is reached with $\phi=1$ (a better global agreement with the exact values for $2 \le N \le 8$ is obtained for $\phi=1.11$). 

\begin{figure*}[htb]
\includegraphics[width=0.48\textwidth]{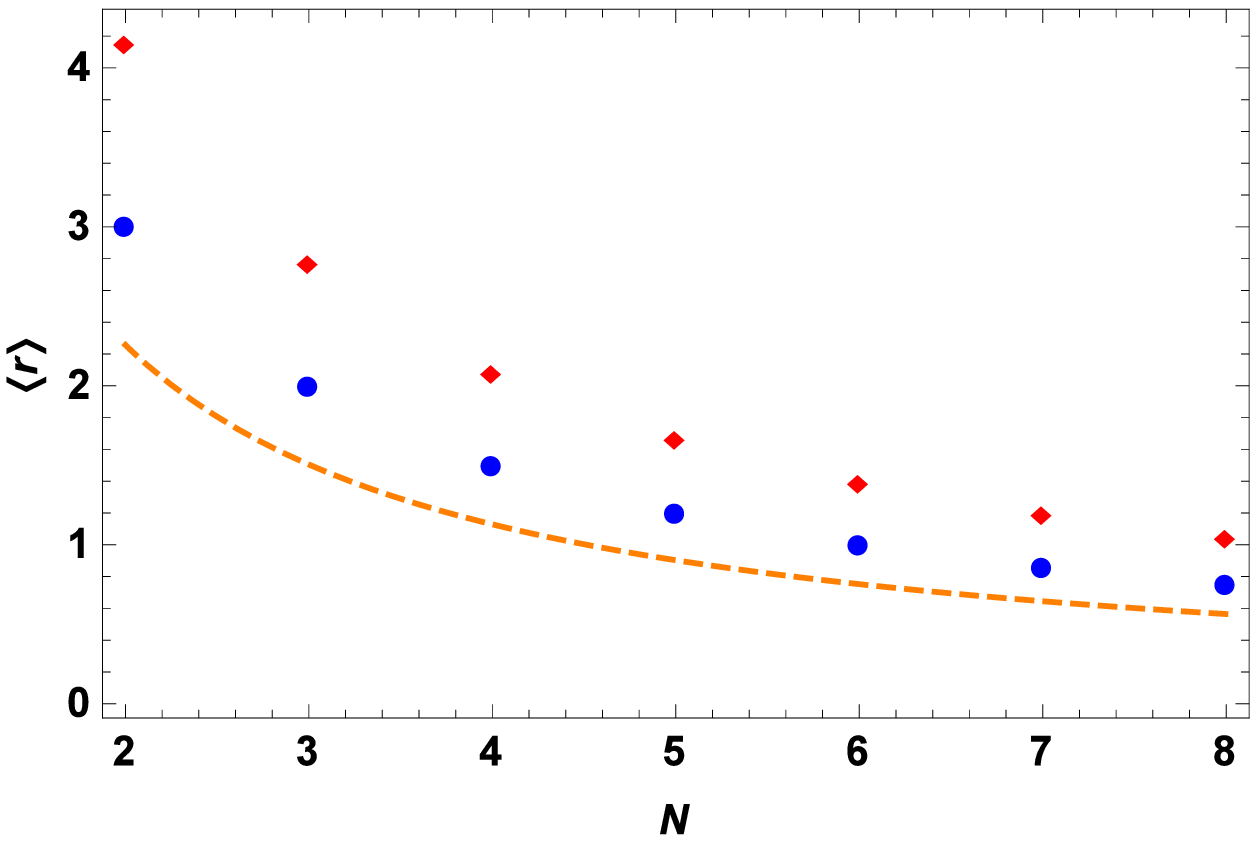} \quad
\includegraphics[width=0.48\textwidth]{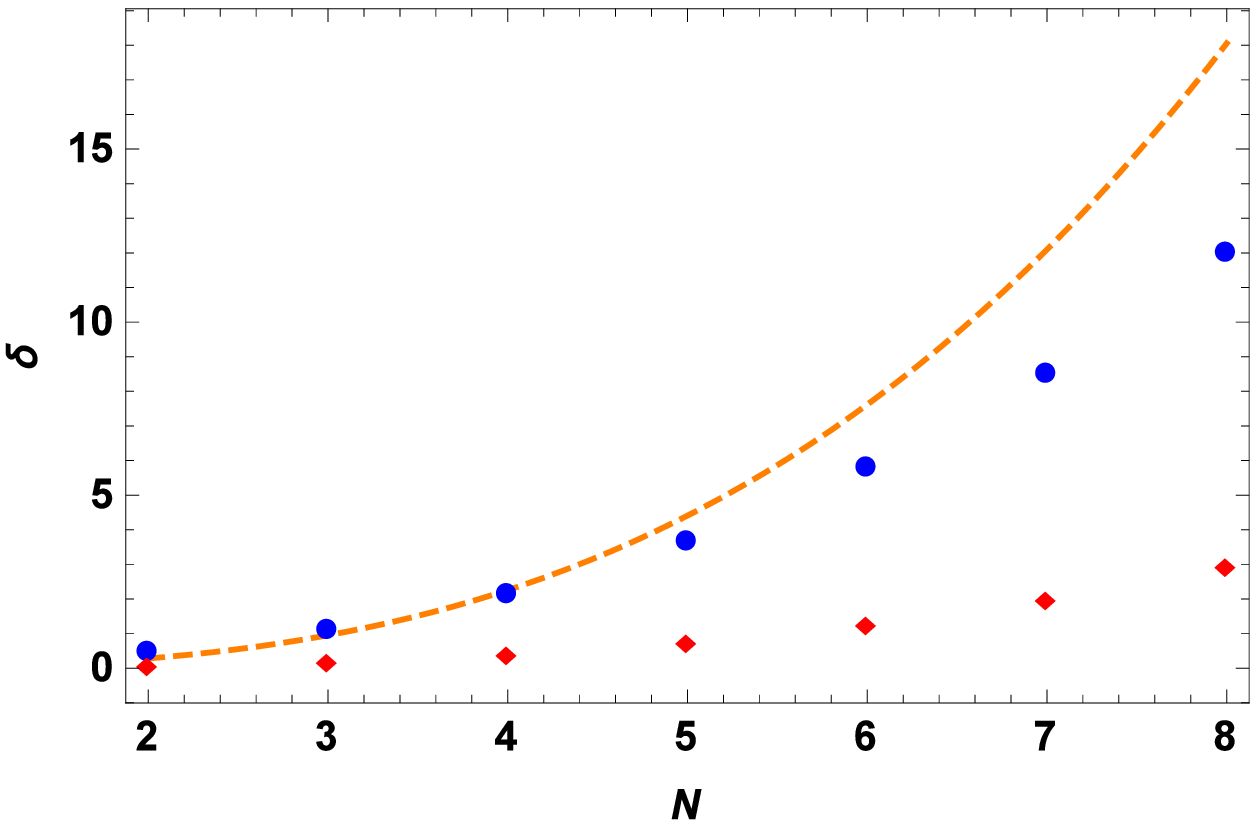}
\caption{$\langle r \rangle$ (left) and $\delta$ (right) for self-gravitating bosons: exact results (circle) \cite{horn14}, ET results for $\phi=2$ (diamond), ET results for $\phi=1$ (dashed line). }
\label{fig:sgbobs} 
\end{figure*}

Mean values $\langle r \rangle$ are presented with $\delta$ in Fig.~\ref{fig:sgbobs}. Again, for the genuine ET, values of $\langle r \rangle$ are globally better reproduced than values of $\delta$. The strong improvement obtained for the energy is not repeated for these observables. 

\subsection{Confined bosons} 
\label{sec:cb}

Now, we consider particles confined by a harmonic oscillator potential with a pairwise repulsive Coulomb interaction
\begin{equation}\label{Hcb}
T(p)=\frac{p^2}{2 m}, \ U(s)=\frac{1}{2}m\,\omega^2 s^2, \ V(r)=\frac{g}{r}.
\end{equation}
The ET method gives the following lower bound
\begin{eqnarray}
\label{Hcbr0}
r_0&=& \frac{N^{5/6}(N-1)^{1/2}}{2^{5/6}} \left(\frac{g}{m\, \omega^2}\right)^{1/3} G_{-}(Y),\\
\label{HcbE}
E&=&\frac{N^{2/3}(N-1)}{2^{5/3}} \left( m\, \omega^2\, g^2\right)^{1/3} \left( G_{-}(Y)^2 + G_{-}(Y)^{-1} \right)\\
\label{HcbY}
&\textrm{with}& Y=\frac{2^{16/3}}{3}\frac{1}{N^{4/3}(N-1)^2} \left(\frac{\omega}{m\, g^2}\right)^{2/3} Q_\phi^2.
\end{eqnarray}
$G_{\pm}(Y)$ is the only positive root of the quartic equation $4 x^4 \pm 8x - 3Y =0$  with $Y \ge 0$ \cite{silv12} \footnote{$G_{\pm}(Y) = \mp \frac{1}{2} \sqrt{V(Y)} + \frac{1}{2} \sqrt{ 4 (V(Y))^{-1/2}- V(Y)}$ with 
$V(Y)=\left(2 + \sqrt{4 + Y^3} \right)^{1/3} -  Y\left(2 + \sqrt{4 + Y^3} \right)^{-1/3}$.}.
One can check that $E=\omega\, Q$ when $g=0$, which is the exact solution in this case.  
 
\begin{figure*}[htb]
\includegraphics[width=0.48\textwidth]{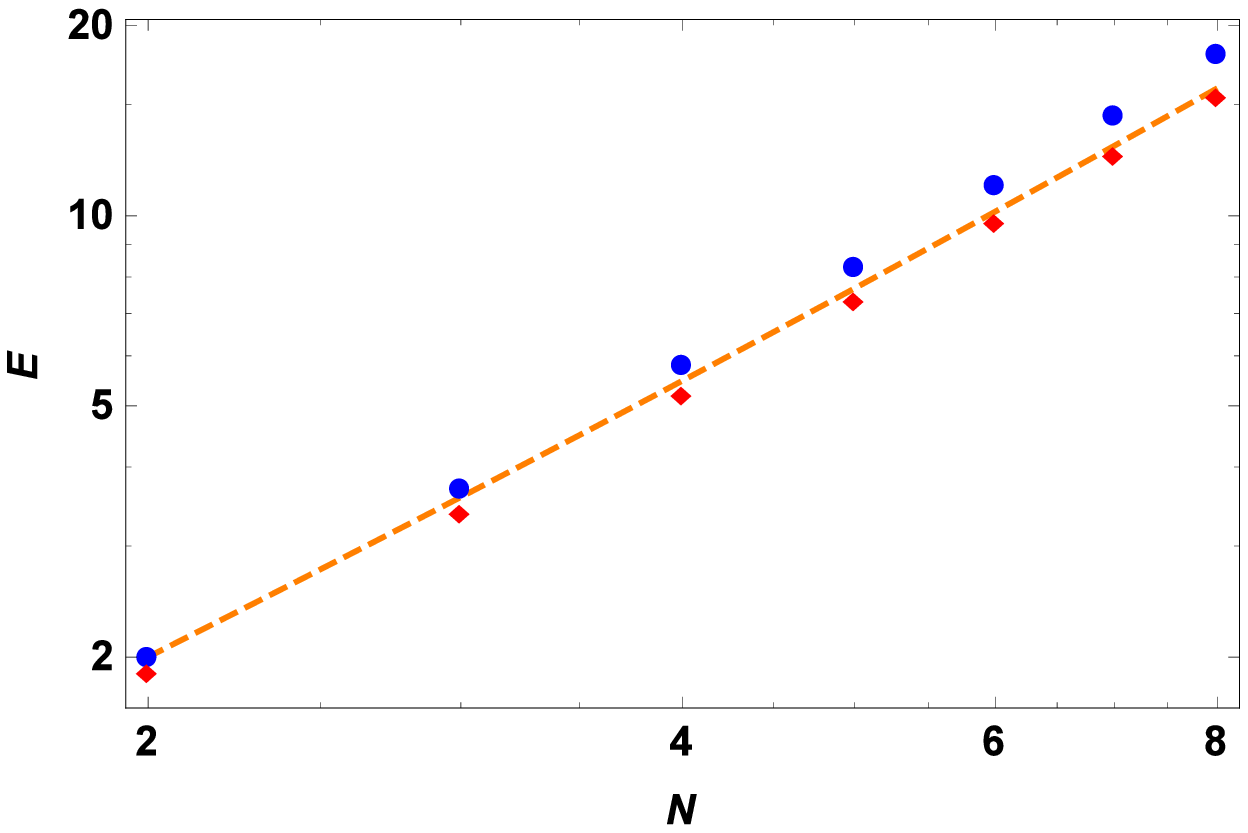} \quad
\includegraphics[width=0.48\textwidth]{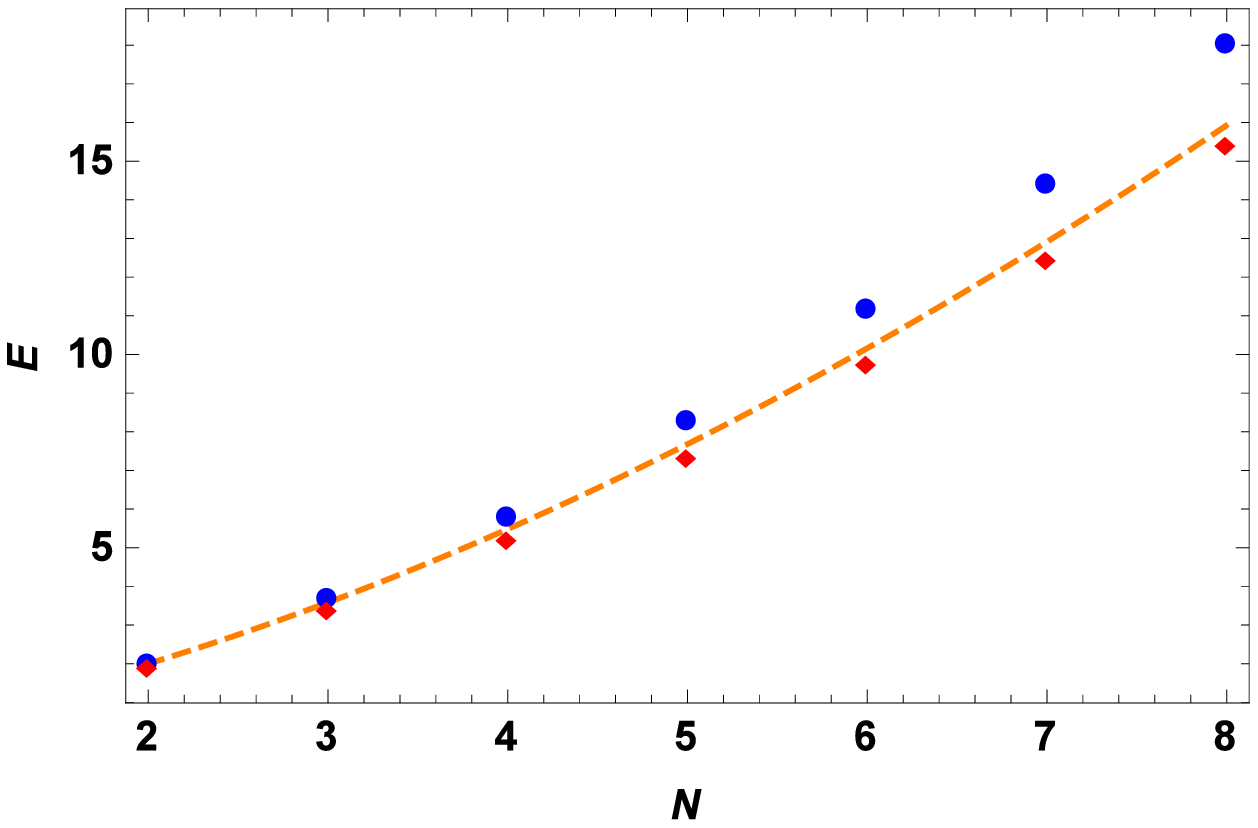}
\caption{Energies, $E$, of confined bosons: exact results (circle) \cite{horn14}, ET results for $\phi=2$ (diamond), ET results for $\phi=2.58$ (dashed line). Left: log-log plot, right: linear plot.}
\label{fig:cbE} 
\end{figure*}

Again, the comparison is performed with \cite{horn14}, for the parameters $m=1$, $\omega=0.5$ and $g=1$. The Hamiltonian studied in this reference is defined with $\sum_{i=1}^N \bm r_i^2=\sum_{i=1}^N (\bm r_i-\bm R)^2+N \bm R^2 $, that is to say with the centre of mass motion ruled by a harmonic oscillator, with a mass $N m$ and a frequency $\omega$. So, to compare the results of this paper with ours, we must add the value $\frac{3}{2}\omega$ (centre of mass ground state energy) to the internal energy of the ground state of our Hamiltonian (\ref{Hcb}). Results for the ground state energies $E$ are presented in Fig.~\ref{fig:cbE}. The agreement is quite good for the genuine ET. With the values chosen for the parameters, an analytical solution can be found for $N=2$ \cite{hall15}. This exact result can be reproduced by our formula with $\phi=2.58$. A better global agreement is obtained for this value of $\phi$, but the improvement is very weak. 

\begin{figure*}[htb]
\includegraphics[width=0.48\textwidth]{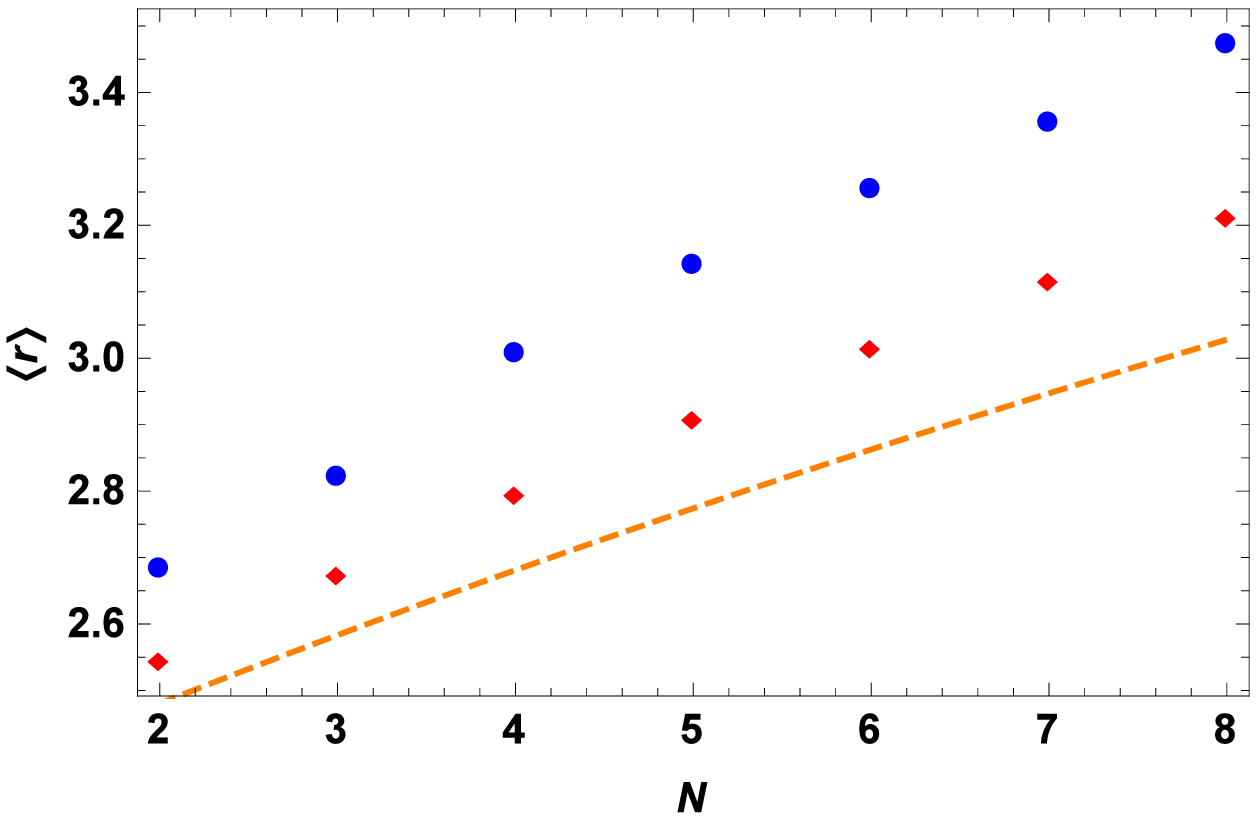} \quad
\includegraphics[width=0.48\textwidth]{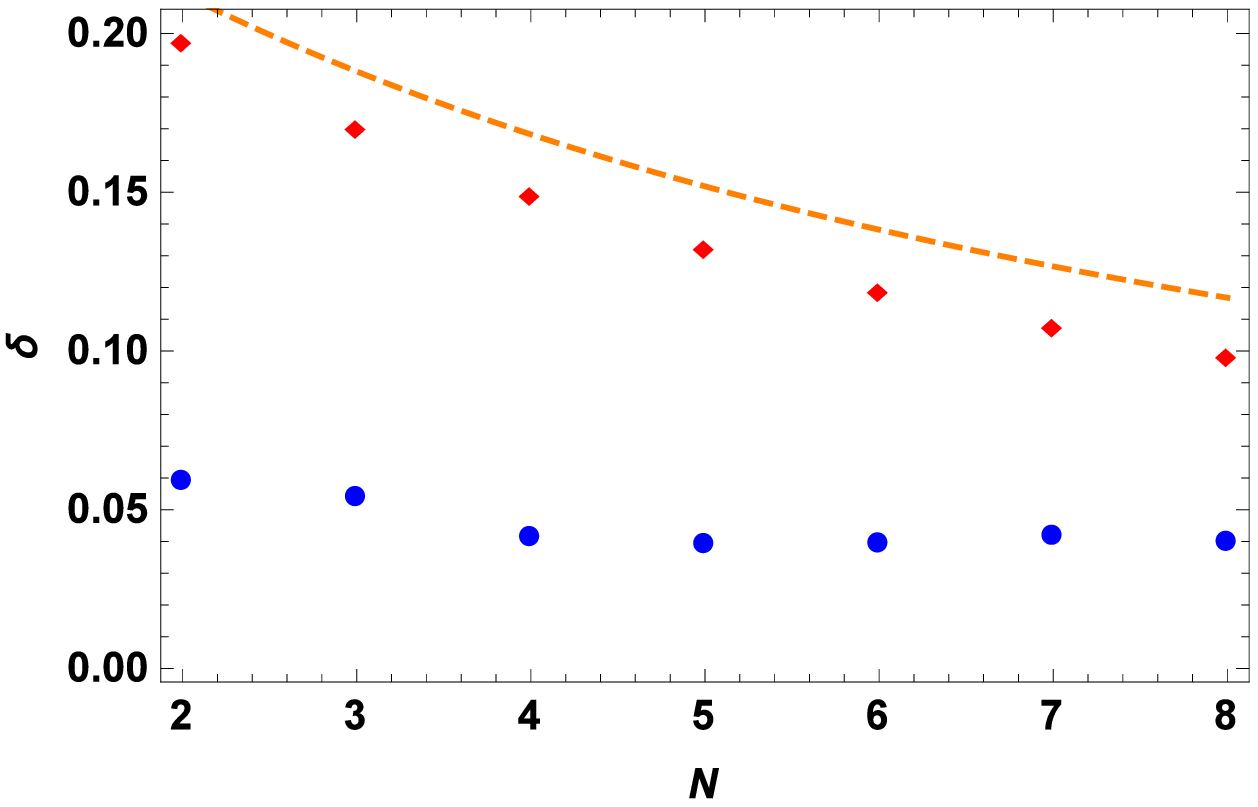}
\caption{$\langle r \rangle$ (left) and $\delta$ (right) for confined bosons: exact results (circle) \cite{horn14}, ET results for $\phi=2$ (diamond), ET results for $\phi=2.58$ (dashed line). }
\label{fig:cbobs} 
\end{figure*}

Mean values $\langle r \rangle$ are presented with $\delta$ in Fig.~\ref{fig:cbobs}. The agreement of the genuine ET results is just reasonable, with again the values of $\langle r \rangle$ better reproduced than the values of $\delta$. In this case, the quality of the observables is deteriorated for $\phi=2.58$.

\subsection{Large-$N$ baryons}
\label{sec:lnb}

Light baryons in a particular large-$N$ limit ($N$ is the number of colours) can be studied with the following Hamiltonian \cite{buis12}
\begin{equation}\label{Hlnb}
T(p)=p, \ U(s)=\lambda\, s, \ V(r)=-\frac{g}{r}.
\end{equation}
The baryons are then composed of $N$ quarks $u$ or $d$, with $m=0$, moving ultrarelativistically. Quarks are fermions, but the colour part of the wave-function is completely antisymmetrical. So, they behave like bosons for the rest of the wave-function. With the ET, the following upper bound is obtained for the mass $E$ of the baryons
\begin{eqnarray}
\label{Hlnbr0}
r_0&=& \frac{1}{\sqrt{\lambda}}\,\sqrt{ N\, Q_\phi - \left(\frac{N(N-1)}{2}\right)^{3/2} g }, \\
\label{HlnbE}
E&=& \sqrt{4\,\lambda}\,\sqrt{ N\, Q_\phi - \left(\frac{N(N-1)}{2}\right)^{3/2} g }.
\end{eqnarray}

In order to test the relevance of $Q_\phi$ for the ground state and some excited states, we consider the 3-body system studied in \cite{silv10}. For the genuine ET ($\phi=2$), the mean relative error $\Delta$ computed on the 16 eigenmasses listed in Table~\ref{tab:1} is 15.1\%. With the value $\phi=\sqrt{2}$ predicted by the DOS method for $g=0$ \cite{sema13b}, $\Delta$ reduces to 4.4\%. With $\phi=1.35$, the ET yields the exact value for the ground states and $\Delta$ drops to 3.1\%. Finally, with $\phi=1.23$, the minimal value of 2.4\% is reached for $\Delta$.

\begin{table}[htp]
\caption{Some eigenmasses in GeV of the Hamiltonian~(\ref{Hlnb}) for $D=3$, $N=3$, $\lambda=0.2$~GeV and $g=\frac{2}{3}\alpha_S$ with $\alpha_S=0.4$  \cite{silv10}. Accurate results obtained from an expansion in a harmonic oscillator basis \cite{silv96} are given in the column ``Exact". Results computed with the ET for four values of $\phi$ are listed in the four last columns, with the associated mean relative errors $\Delta$. The sums $n_1+n_2$ and $l_1+l_2$ are the quantum numbers of the main component of the corresponding genuine eigenstate in the harmonic oscillator basis. These numbers are used for the computation of $Q_\phi$.}
\begin{center}
\label{tab:1}
\begin{tabular}{lllllll}
\hline\noalign{\smallskip}
 & & & \multicolumn{4}{c}{ET} \\
$n_1+n_2$ & $l_1+l_2$ & Exact & $\phi=2$ & $\phi=\sqrt{2}$ & $\phi=1.35$ & $\phi=1.23$ \\
\noalign{\smallskip}\hline\noalign{\smallskip}
0 & 0 & 2.128 & 2.468 & 2.165 & 2.128 & 2.060 \\
0 & 1 & 2.606 & 2.914 & 2.662 & 2.633 & 2.578 \\
1 & 0 & 2.739 & 3.300 & 2.842 & 2.788 & 2.682 \\
0 & 2 & 2.959 & 3.300 & 3.080 & 3.055 & 3.007 \\
1 & 1 & 3.125 & 3.646 & 3.237 & 3.189 & 3.098 \\
0 & 3 & 3.299 & 3.646 & 3.448 & 3.425 & 3.383 \\
2 & 0 & 3.260 & 3.961 & 3.387 & 3.318 & 3.186 \\
1 & 2 & 3.422 & 3.961 & 3.589 & 3.546 & 3.463 \\
0 & 4 & 3.581 & 3.961 & 3.780 & 3.759 & 3.721 \\
2 & 1 & 3.584 & 4.253 & 3.725 & 3.662 & 3.542 \\
1 & 3 & 3.716 & 4.253 & 3.909 & 3.869 & 3.794 \\
0 & 5 & 3.861 & 4.253 & 4.085 & 4.066 & 4.030 \\
3 & 0 & 3.721 & 4.527 & 3.856 & 3.775 & 3.619 \\
2 & 2 & 3.838 & 4.527 & 4.034 & 3.976 & 3.866 \\
1 & 4 & 3.966 & 4.527 & 4.205 & 4.168 & 4.098 \\
0 & 6 & 4.103 & 4.527 & 4.369 & 4.351 & 4.318 \\
\noalign{\smallskip}\hline\noalign{\smallskip}
$\Delta$ & & & 15.1\% & 4.4\% & 3.1\% & 2.4\% \\
\noalign{\smallskip}\hline
\end{tabular}
\end{center}
\end{table}

\section{Conclusion}
\label{sec:conclu}

The envelope theory is a powerful method to compute eigenvalues for quite general $N$-body systems with identical particles in $D$ dimensions \cite{sema13a}. The method is easy to implement since it reduces to find the solution of a transcendental equation. Its interest is not to produce accurate results, but to yield rapidly reliable estimation of energies and observables. Here, the method is tested with the ground states of three nonrelativistic systems, weakly interacting bosons, self-gravitating bosons, and confined bosons, accurately computed in \cite{horn14} up to $N=8$. A fourth case is an ultrarelativistic system of quarks (boson-like) studied in \cite{silv10}. For all these cases, an analytical variational bound is computed. Considering the simplicity of the method, quite good results can be obtained for energies and observables in all the systems considered. 

A universal effective quantum number for centrally symmetric 2-body systems is proposed in \cite{loba09}. As it seems relevant to extend this notion to 3-body systems \cite{sema13b}, the characteristic global quantum number of the envelope theory is modified accordingly. Only one parameter must be fixed either by theoretical considerations or by a fit on a single known accurate solution. The variational character of the energies cannot then be guaranteed, but the improvement is generally good, even for the excited states. Unfortunately, the behaviour of observables is not really predictable. 

Since $N$-body problems are always difficult and heavy to solve accurately, the envelope theory can be used as a guide for the study of these complicated systems. It could be interesting to test the method with other systems in various dimensions, for different values of $N$, and not only for the ground state. In particular, it is not certain that the modification of the global quantum number proposed in this work can always lead to improvement of the energies.

\end{document}